\documentclass[11pt,a4paper,showkeys,showpacs]{article}
\usepackage{alpha}
\usepackage{defs}
\usepackage{cleveref}
\usepackage{booktabs}
\usepackage{longtable}
\usepackage{color, colortbl}

\begin{document}

\title{
\begin{flushright}
\vspace{-2cm}
 \small{
CERN-TH-2023-230\\
WUB/23-02\\
HU-EP-23/72\\
DESY-24-001 
}
\vskip 0.7cm
\end{flushright}
Heavy Wilson Quarks and O($a$) Improvement: \\[0ex] Nonperturbative Results for $\bg$\\[2ex]
\href{https://www-zeuthen.desy.de/alpha/}{\includegraphics[width=2.8cm]{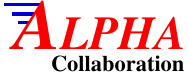}}}
\date{\today}

\author{\vspace{-1cm}Mattia~Dalla~Brida}
\address{\vspace{-4mm}Theoretical Physics Department, CERN, 1211 Geneva 23, Switzerland}
\author{Roman H{\"o}llwieser, Francesco Knechtli, Tomasz Korzec}
\address{\vspace{-5mm}Deptartment of Physics, University of Wuppertal, Gau{\ss}strasse 20, 42119 Germany}
\author{Stefan~Sint}
\address{\vspace{-5mm}School of Mathematics and Hamilton Mathematics Institute, Trinity College Dublin, Dublin 2, Ireland}
\author{Rainer~Sommer}
\address{\vspace{-4mm}Deutsches Elektronen-Synchrotron DESY, Platanenallee~6, 15738~Zeuthen, Germany\\
Institut~f\"ur~Physik, Humboldt-Universit\"at~zu~Berlin, Newtonstr.~15, 12489~Berlin, Germany\vspace{-1.5cm}}

\begin{abstract}
With Wilson quarks, on-shell O($a$) improvement of the lattice QCD action is achieved by including
the Sheikholeslami-Wohlert term and two further operators of mass dimension 5, 
which amount to a mass-dependent  rescaling of the bare parameters. We here focus on the rescaled bare coupling,
$\g02t = g_0^2(1+\bg a\mq)$, and the determination of $\bg(g_0^2)$, which is currently only known
to 1-loop order of perturbation theory. We derive suitable improvement conditions in the chiral limit and
in a finite space-time volume and evaluate these for different gluonic observables, both with and without the gradient flow.
The choice of $\beta$-values and the line of constant physics are motivated by the ALPHA collaboration's 
decoupling strategy to determine $\alpha_s(m_Z)$~\cite{DallaBrida:2022eua}.  However, 
the improvement conditions and some insight into systematic effects may prove useful in other contexts,~too.
\end{abstract}

\maketitle

\tableofcontents

\newpage

\section{Introduction}
\label{sec:intro}

On-shell O($a$) improvement of lattice QCD with  Wilson quarks near the chiral limit 
requires a single new term in the action, the Sheikholeslami-Wohlert term with coefficient $\csw$~\cite{Sheikholeslami:1985ij}. 
The latter is a function of the bare coupling $g_0^2$ and can be determined by imposing continuum chiral symmetry at
finite lattice spacing~\cite{Luscher:1996sc}. With $\Nf$ degenerate
quarks of subtracted bare mass $\mq= m_0-\mcr$, there are 2 further counterterms 
which can be implemented as a rescaling of the bare parameters, i.e.
\begin{equation}
     \g02t= g_0^2(1+\bg(g_0^2) a\mq), \qquad \mqt = \mq(1+\bm(g_0^2)a\mq)\,.
\end{equation}
In practice, perturbative estimates of these $b$-coefficients are often sufficient,
as $a\mq$ is a small parameter for the light quark flavours up, down and strange. For the most
common gauge actions, i.e.~Wilson's plaquette or the tree-level O($a^2$) improved L\"uscher-Weisz actions,
the one-loop result for $\bg$ reads~\cite{Sint:1995ch},
\begin{equation}
   \bg(g_0^2) = 0.01200\times \Nf g_0^2 + \rmO(g_0^4).
   \label{eq:bgpert}
\end{equation}
For heavier quarks, $a\mq$ is no longer small, and in the recent decoupling project
of the ALPHA collaboration, values of $a\mq$ up to $0.3-0.4$ are included in the analysis~\cite{DallaBrida:2019mqg,DallaBrida:2022eua}.
In such situations a non-perturbative determination of the relevant $b$-coefficients
becomes very desirable as, otherwise, uncontrolled systematic errors may ensue. In particular,
the decoupling result for $\alpha_s(m_Z)$~\cite{DallaBrida:2022eua} includes a sizeable systematic error due to 
an assumed 100 percent uncertainty on the truncated perturbative result for $\bg$, Eq.~(\ref{eq:bgpert}).

Various $b$-coefficients have been determined non-perturbatively over the past 25 years, including $\bm$
and some of the coefficients required for the O($a$) improvement of quark bilinear composite operators
(see~\cite{deDivitiis:1997ka,Guagnelli:2000jw,Bhattacharya:2000pn,Bhattacharya:2005ss,Fritzsch:2010aw,Korcyl:2016ugy,Fritzsch:2018zym,
Gerardin:2018kpy,deDivitiis:2019xla} for an incomplete list of references).
The coefficient $\bg$ is a notable exception. While there exist some qualitative ideas in the literature
as to how $\bg$ could be determined by measurement~(see e.g.~\cite{Martinelli:1997zc}), no proof of concept regarding 
their viability and practicality has been given. 
This is somewhat surprising given the central r\^ole of $\bg$ for the consistency of O($a$) improvement: 
it is required when the quark masses are varied at fixed lattice spacing such as needed for chiral extrapolations
or in studies of decoupling~\cite{DallaBrida:2019mqg,DallaBrida:2022eua}. Furthermore, since constant lattice spacing amounts 
to keeping $\g02t$ constant, the renormalization constants in mass-independent schemes 
depend on $\g02t$ rather than $g_0^2$. In fact, for composite operators this often leads to the determination 
of effective $b$-coefficients, which include the $\bg$-contribution from the renormalization
constant by a Taylor expansion to first order in $a\mq$ \cite{Korcyl:2016ugy}, 
even if not mentioned explicitly~\cite{Fritzsch:2018zym}.

In this paper we propose a class of improvement conditions to determine $\bg$ non-perturbatively and then
proceed to evaluate these for $\Nf=3$, non-perturbatively O($a$) improved lattice QCD~\cite{Bulava:2013cta} with
the tree-level O($a^2$) improved L\"uscher-Weisz gauge action~\cite{Luscher:1984xn}.
We perform  consistency checks by looking at gradient flow observables as well as Creutz ratios in a finite space time
volume and for  different lines of constant physics. We also generated some data in the small coupling region in order to
compare to perturbation theory. Our parameter choices, in particular the range of lattice spacings, 
are motivated by the decoupling project~\cite{DallaBrida:2019mqg,DallaBrida:2022eua} and do not overlap with the parameters of 
CLS~\cite{Bruno:2014jqa,cls:status}.

The paper is organized as follows. In Section~\ref{sec:conditions} we discuss improvement conditions for $\bg$. In particular
we review the  connection to the O($a$) improvement term in the flavour singlet scalar density, which clarifies and corrects
the discussion in~\cite{Bhattacharya:2005rb}. We then define our set of observables (Section~\ref{sec:observables}) 
and present the chosen line of constant physics and the corresponding simulation parameters (Section~\ref{sec:lcp}). 
In Section~\ref{sec:data}, we discuss some details of the data analysis and various consistency checks before we present our results for $\bg$ (Section~\ref{sec:results}) 
and our conclusions (Section~\ref{sec:conclusions}). A few technical details have been delegated to appendices: Appendix~\ref{app:A} gives some details
pertaining to the derivation of the key equations in Section~\ref{sec:conditions}. Appendix~\ref{app:B} contains some additional information on the simulations and corresponding parameters,
while the simulation results are tabulated in Appendix~\ref{app:C}.

\section{Improvement conditions for $\bg$}
\label{sec:conditions}
%
\subsection{Chiral Ward identities and $\bg$}

While improvement coefficients such as $\csw$ or $\ca$, multiplying the O($a$) counterterms 
in the action and to the axial current, respectively, can be
determined by chiral Ward identities~\cite{Luscher:1996sc}, this is less obvious for the $b$-coefficients
multiplying $a\mq$. One problem is that, away from the chiral limit, the on-shell condition is potentially violated
by unavoidable contact terms, at least for chiral Ward identities beyond the simple PCAC relation.
A systematic analysis for the $\Nf=3$ lattice QCD action and all quark bilinear composite fields 
has been carried out by Bhattacharya et al.~in \cite{Bhattacharya:2005rb}. By studying the general case
of mass non-degenerate quarks, they show how violations of the on-shell condition can be by-passed
and they list all combinations of renormalization constants and improvement coefficients
which are, at least in principle, determined by chiral Ward identities.
The coefficient $\bg$ is included in this list, through its relation to the O($a$) counterterm 
of the flavour singlet scalar density, $S^0(x) = \psibar(x)\psi(x)$.
In the mass-degenerate limit, the corresponding renormalized and O($a$) improved operator 
takes the form~\cite{Sint:1999ke,Bhattacharya:2005rb}, 
\begin{equation}
  S^{0}_{\rm R}  =  Z_{{\rm S}^0} \left(1+b_{{\rm S}^0}a\mq\right)
  \left( {S}^0  + c_{\rm S} a^{-3} + d_{\rm S} a \tr\left\{F_{\mu\nu} F_{\mu\nu}\right\}\right)\,. 
  \label{eq:SRstructure}
\end{equation}
Using an argument based on Feynman-Hellmann type identities, Bhattacharya et al.~derive the 
identity\footnote{In the notation of~\cite{Bhattacharya:2005rb}, $g_{\rm S}=-d_{\rm S}$, due to
their convention of hermitian, rather than anti-hermitian $F_{\mu\nu}$.}
\begin{equation}
  \bg =-2g_0^2 d_{\rm S}\,,
  \label{eq:bg-ds}
\end{equation}
and $d_{\rm S}$ can, in prinicple, be determined by chiral Ward identities, 
together with $c_{\rm S}$ and the scale independent ratio $Z_{\rm P}/Z_{\rm S^0}$~\cite{Sint:1999ke}.
A difficulty is posed by the fact that the chiral limit is only defined up to cutoff effects of O($a^2$), 
which render the subtraction of the cubic power divergence ambiguous. 
More precisely, with $c_{\rm S} \equiv c_{\rm S}(g_0^2,a\mq)$, an expansion in powers of $a\mq$ yields a 
term $\propto a\mq / a^3 = \mq/a^2$. 
Given the intrinsic O($a^2$) ambiguity of $\mq$, this results in an ambiguity of O($1$) in the subtraction term
rendering the definition of the renormalized scalar density impossible along these lines. 
It is therefore important to note that one may combine Ward identities such that
only connected correlation functions remain, in which all power divergences cancel. 
A practical implementation using SF boundary conditions~\cite{Luscher:1992an,Sint:1993un} can be obtained 
following the discussion in~\cite{Sint:1999ke}. Whether these observations lead to a practical method for a non-perturbative 
determination of $\bg$ remains to be seen.
We emphasize that this strategy relies entirely on the identity~(\ref{eq:bg-ds}). We would therefore like to point out a subtlety 
that was overlooked in~\cite{Bhattacharya:2005rb}. Given that one relates an improved operator to the improved action parameters, it is natural
to expect that the O($a$) counterterm in Eq.~(\ref{eq:SRstructure}) cannot be discretized arbitrarily, but must be related 
to the derivative of the lattice action density with respect to $g_0^2$. For definiteness, the O($a$) improved lattice action
\begin{equation}
  S = S_g + a^4 \sum_x \psibar(x) \left(D_{\rm W} + m_0 +\csw\frac{ia}4 \sigma_{\mu\nu}F_{\mu\nu}(x)\right)\psi(x),
\label{eq:lataction}
\end{equation} 
includes the lattice gauge action $S_g = a^4\sum_x {\cal L}_g(x)$ with ${\cal L}_g$ being
the lattice version of the (Euclidean) Yang-Mills Lagrangian density in the continuum, 
\begin{equation}
   {\cal L}_{\rm YM} = -\dfrac{1}{2g_0^2}\tr\left\{ F_{\mu\nu}(x) F_{\mu\nu}(x)\right\}.
   \label{eq:YMLagrange}
\end{equation}
We find that, due to the $g_0$-dependence of $\csw$, a contribution from the fermionic part of the action must be included,
so that the correct lattice discretization of the counterterm is given by,
\begin{equation}
   \tr\left\{F_{\mu\nu} F_{\mu\nu}\right\} \quad \rightarrow \quad
   -2g_0^2 \left({\cal L}_g - g_0^2 \times \frac{ia}{4}\csw'\psibar \sigma_{\mu\nu}F_{\mu\nu}\psi\right),
   \label{eq:ds-counterterm}
\end{equation}
with $\csw' \equiv \frac{d}{dg_0^2}\csw$. For future reference we provide a detailed derivation in Appendix~\ref{app:A}. 

\subsection{Restoration of chiral symmetry in a small volume}

In this paper we pursue a different approach. It is based on the observation that, 
in the absence of spontaneous chiral symmetry breaking, physical quark mass effects in gluonic observables 
are quadratic or higher order in  the quark mass. In contrast, the counterterm proportional 
to $\bg$ is designed to cancel terms that are linear in the quark mass, allowing us, in principle, 
to distinguish these effects.
Let us first have a closer look at the physical quark mass dependence. If we consider a finite space time manifold
without boundaries, the lattice QCD partition function with Ginsparg-Wilson
type quarks~\cite{Ginsparg:1981bj,Hasenfratz:1998jp,Neuberger:1997fp,Luscher:1998pqa,Luscher:1998kn} becomes a 
finite dimensional, mathematically well-defined ordinary and Grassmann-integral, with exact
chiral and flavour symmetries. The absence of spontaneous symmetry breaking in a finite volume,
implies that the partition function is an analytic function of $\mq$. For even $\Nf$, 
a change of the fermionic variables by a discrete chiral field transformation\footnote{On the lattice with Ginsparg-Wilson quarks and 
Neuberger's ($\gamma_5$-hermitian) Dirac operator $D_N$ \cite{Neuberger:1997fp}, the first of the $\gamma_5$ factors is replaced by 
$\hat{\gamma}_5\equiv\gamma_5(1-aD_N)$~\cite{Luscher:1998kn}. 
Note that, for even $\Nf$, the transformation is part of the {\em non-singlet} chiral symmetry.}
\begin{equation}
  \psi \rightarrow \gamma_5 \psi,\qquad \psibar \rightarrow -\psibar \gamma_5\,,
\end{equation}
then establishes that this function must be even, implying the absence of 
terms linear in $\mq$ for any gluonic observable.
With odd $\Nf>2$, this is no longer true; however, by adapting the argument
of appendix~D in ref.~\cite{DallaBrida:2020gux} to the lattice regularization with 
Ginsparg-Wilson quarks, the discrete chiral field transformation,
\begin{equation}
    \psi \rightarrow \exp\left(i\frac{\pi}{\Nf}\gamma_5\right) \psi,\qquad 
    \psibar \rightarrow \psibar \exp\left(i\frac{\pi}{\Nf}\gamma_5\right)\,,
\end{equation}
leads to a change of variables with unit Jacobian that leaves the massless action invariant. Under the same transformation, a single insertion of the flavour singlet scalar density, $S^0$, into a gluonic correlation function is proportional to itself and must vanish,  provided that both the QCD action and the gluonic observable are parity even.\footnote{We again use a continuum notation; on the lattice,
besides using $\hat{\gamma}_5$ \cite{Luscher:1998kn} for the transformation of $\psi$, the scalar lattice density takes the form 
$S^0= \psibar(1-\frac{a}{2}D_N)\psi$, such that it correctly transforms under the lattice chiral symmetry 
(see e.g.~Section~2 in \cite{Frezzotti:2000nk}).}
Although this does not exclude all odd powers in the quark mass, it means that
contributions linear in the quark mass are indeed  absent.

We will denote a generic renormalized gluonic observable by $O_g$. Gradient flow observables~\cite{Narayanan:2006rf,Luscher:2010iy}, 
such as the gauge action density at finite flow time~\cite{Luscher:2010iy}, are natural candidates, 
but the gradient flow is not a necessary requirement. We choose a finite Euclidean space time volume $L^4$ 
with boundary conditions that do not break all the chiral symmetries. A hyper-torus geometry with some kind of
periodic boundary conditions for all fields is an obvious possibility, however, other possible options include
chirally rotated SF ($\chi$SF) boundary conditions  (with even $\Nf$)~\cite{Sint:2010eh,DallaBrida:2016smt}, 
or a mixture of SF~\cite{Luscher:1992an,Sint:1993un} and $\chi$SF boundary conditions (with odd $\Nf$)~\cite{DallaBrida:2018tpn}. 
For simplicity we also assume that $O_g$ has no mass dimension, which can always be achieved 
by multiplication with the appropriate
power of $L$.  In the continuum limit, the observable thus becomes a function of two dimensionless variables, which
we may choose as $z=ML$ and $\Lambda_\msbar^{(3)}L$, with $M$ and $\Lambda_\msbar^{(3)}\equiv \Lambda$ being the Renormalization Group
Invariant (RGI) quark mass and $\Lambda$-parameter of the $\Nf=3$ theory, respectively.
From the previous discussion we then  expect that the continuum limit of the lattice expectation value, 
$\langle O_g\rangle$, becomes a function of  $(z,\Lambda L)$, with an expansion in the RGI mass of the form
\begin{equation}
   \langle O_g\rangle = \langle O_g\rangle_{z=0} + A(\Lambda L) z^2 + \rmO(z^3)\,. 
\end{equation}
When considering this observable on the lattice, the mass dependence will have a linear term in $z$, unless $\bg$ is chosen correctly.
Hence, a possible improvement condition for $\bg$ is given by
\begin{equation}
  \left.\dfrac{\partial\langle O_g\rangle }{\partial z}\right\vert_{z=0,\text{$L\Lambda =$ const}} = 0\,. 
\end{equation} 
In order to derive an explicit equation for $\bg$, we will proceed in two steps. First we keep the lattice size $L/a$ and spacing $a$ fixed.
Up to O($a^2$) effects this can be achieved by working at fixed $\g02t$. The choice for a line of constant physics only matters later,
when $\g02t$ and thus the lattice spacing is changed. This will be discussed in Section~\ref{sec:lcp}.
Since $z$ is proportional to $\mqt$, we may change variables from $(z, \Lambda L,L/a)$ to $(a\mqt,\g02t,L/a)$ and the improvement condition
at a given lattice spacing then reads
\begin{equation}
   \left.\dfrac{\partial\langle O_g\rangle }{\partial a\mqt}\right\vert_{\mqt=0,\g02t} = 0\,. 
\end{equation} 
We would now like to perform a final change of variables from improved to unimproved bare parameters. In fact,
it is technically convenient to do the reverse transformation and then solve a $2\times 2$ linear system.
For the sake of readability we have relegated this discussion to Appendix~\ref{app:A}. The improvement condition 
then implies,
\begin{equation}
  \bg(g_0^2)=\left.\dfrac{\partial \langle O_g \rangle}{\partial a\mq}\right\vert_{g_0^2,\mq=0} 
    \times \left[ \left. g_0^2 \dfrac{\partial \langle O_g \rangle}{\partial g_0^2} \right\vert_{\mq=0} \right]^{-1}\,.
  \label{eq:bg}
\end{equation}
This is the desired equation for $\bg$ in terms of the bare parameters, which are the ones directly controlled in numerical simulations.
Note that the condition is again formulated in the chiral limit, $\mq=0$. 

At this point we recall that, with Wilson quarks, the massless limit is not unambiguously defined. 
What is required here is a definition up to an O($a^2$) ambiguity, which can be achieved
by tuning the improvement coefficients $\csw$ in the action and $\ca$ in the improved
axial current used for the definition of the PCAC mass~\cite{Luscher:1996sc}. 
Note that any remnant O($a$) ambiguity of the chiral limit would make the quadratic physical mass effect 
vanish only up to O($a$) corrections,  which, by virtue of the $1/a$ factor in the $a\mq$-derivative 
would contribute at O($1$), leading to a wrong result for the $\bg$-estimate.

A number of choices must still be made. First, one needs to pick suitable gluonic observables, $O_g$. 
Second, one needs to find a practical way to implement the derivatives with respect to the bare parameters
and then one needs to follow a line of constant physics as $\beta=6/g_0^2$ and thus the lattice spacing is changed.

\section{Choice of observables}
\label{sec:observables}

We choose a space-time volume with linear extent $L$ in all directions and a hyper-torus topology.
We use periodic boundary conditions for the gauge fields, and antiperiodic boundary conditions for the fermions
in all 4 space-time directions. The absence of a boundary avoids potential violations of chiral symmetry, while 
antiperiodic boundary conditions for the fermions allow us to perform simulations around the chiral limit (cf.~Section~\ref{sec:simulations} and Appendix \ref{app:B}).
To finish the description of our set-up it remains to specify our choices of gluonic observables, $O_g$.

\subsection{Gradient flow energy density}
\label{sec:bgGF}

First, we consider the action density in terms of
the gradient flow field tensor $G_{\mu\nu}(t,x)$.  
In a continuum language it is given by
\begin{equation}
  \sigma(c) = \sum_{\mu,\nu = 0}^3
    \dfrac{t^2\,\langle {\rm tr}\left\{G_{\mu\nu}(t,x)G_{\mu\nu}(t,x)\right\}\delta_{Q(t),0} 
    \rangle}{\langle \delta_{Q(t),0}\rangle}\Bigg|_{8t=c^2L^2}\,.
 \label{eq:GF}
\end{equation}
In the numerical evaluation translation invariance w.r.t.~$x$ is
of course used. The projection to the sector of vanishing topological charge $Q(t)$ is performed to avoid large autocorrelation times as discussed in \cite{Fritzsch:2013yxa,DallaBrida:2016kgh}. While the motivation is algorithmic, it is part of the definition of
the observable. The gradient flow observable is finite: proven to all orders of perturbation theory~\cite{Luscher:2011bx} and well established in numerical simulations.  
The remaining parameter $c$ fixes the ratio between the scale set by the flow time and the box length~\cite{Fodor:2012td}. On the lattice we discretize the gradient flow equations through the 
$\rmO(a^2)$ improved Zeuthen flow definition, while the action density of the flow fields
is obtained from an $\rmO(a^2)$ improved combination of plaquette and clover
discretizations (see~\cite{Ramos:2015baa,DallaBrida:2016kgh} for the details).

\subsection{Creutz ratios}
To check for $\rmO(a)$ ambiguities in $\bg$ we also use a second observable. It is very independent from $\sigma$, namely it is defined without the gradient flow. We start from rectangular
Wilson loops $W(R,T)$  and form Creutz ratios~\cite{Creutz:1980hb}
\begin{equation}
	\chi(R,T)=-\tilde\partial_R \tilde\partial_T \log(W(R,T))\,,\label{eq:chi1}
\end{equation}
where the lattice derivatives $\tilde\partial_y f(y)=[f(y+a)-f(y-a)]/(2a)$
are symmetric to avoid linear terms in $a$. Note that, analogously to the energy density at positive flow time in eq.~(\ref{eq:GF}), also the Wilson loops $W(R,T)$
defining the Creutz ratios are projected to the topologically trivial sector.
These Creutz ratios are made dimensionless and restricted to the diagonal ones, 
\begin{equation}
   \hat\chi = \left [R^2\,\chi(R,R)\right]_{R=L/4} \label{eq:chi2}\,.	
\end{equation}
We have fixed the ratio $R/L$ to $1/4$ since this is a practical number for lattices with $L/a=12,16,\ldots$, which we
will use below. Also $\hat\chi$ is a finite pure gauge observable
for the following reasons. In continuous space-time, Wilson loops
around smooth paths have been shown to be finite up to an
overall factor $Z_W(l)$ which depends (apart from coupling and regularisation parameter) only on the length $l$ of the path \cite{PhysRevD.24.879}. When the path contains cusps, such as the rectangular Wilson loops, there is an additional renormalization factor depending on the angles
of the cusps. Both factors are removed by the derivatives
in \cref{eq:chi1}. In our implementation on the lattice, we
apply one HYP2 smearing step with parameters $\alpha_1=1,\,\alpha_2=1$ and $\alpha_3=0.5$ to the link variables which form the loop~\cite{Hasenfratz:2001hp, DellaMorte:2003mw, DellaMorte:2005nwx, Grimbach:2008uy, Donnellan:2010mx}. Since this does not change the symmetry properties of the links,
$\hat\chi$ remains finite and has the same continuum limit as the
observable with no smearing. Finally we need also the absence 
of terms linear in $a$ in the observable. Their absence for the static quark potential has been
explained in \cite{Necco:2001xg}. Analogously it  holds for Creutz ratios defined with symmetric derivatives.

\section{Line of constant physics (LCP) and simulations}
\label{sec:lcp}

\subsection{LCP}

\label{subsec:lcp}
The Symanzik expansion applies when the lattice spacing $a$ is changed with all other scales being fixed. Consequently also
improvement conditions need to be enforced along a LCP. 
Since vanishing quark masses are required for our improvement 
condition, it 
only remains to fix $L$ in physical units.%
\footnote{The values, $\kappa_c$, of the hopping parameter where the quark mass vanishes
are taken from a fit similar to the one described in appendix A.1.4 of~\cite{DallaBrida:2016kgh}.
The data are the same, but instead of separate fits at each value of $L/a$, we performed a global fit 
in which the lattice artifacts are parameterized by $(a/L)^3$ times a power series in $g_0^2$. We include the six
lowest powers and such a fit describes the data well, is in good agreement with the previous fit in the 
range of $\beta$ and $L/a$ values where we have data, and most importantly allows for inter/extrapolations in $L/a$.
We have
checked that the uncertainties of $\kappa_c$ are negligible in our analysis. This statement refers to the statistical precision as well as to systematics coming from
data interpolation and extrapolation to large values of $L/a$ at
the larger $\beta$'s.
}
To this end we choose $\gbarGF^2(L)=3.949$, where $\gbarGF^2(L)$ is the gradient flow coupling defined in a finite volume 
with spatial extent $L$ and SF boundary conditions (cf.~\cite{DallaBrida:2022eua} for a precise definition).
For resolutions $L/a\in \{12,16,20,24 \}$ the required $\beta$-values are found in
Table 13 of \cite{DallaBrida:2022eua} with sufficient precision.
They are in the range $\beta\in [4.302,4.7141] $.
For larger $L/a$, it turned out that a precise computation of 
$\bg$ becomes prohibitively expensive.
We therefore deviate from the strict LCP and set $L/a=24$ for 
$ \beta\geq 4.9$. At that point, the lattice spacing in units 
of the $\Lambda$-parameter is very small and also the only additional 
scales determining $a$-ambiguities, $\sqrt{8t}$ as well as $R$,
are large enough to make these ambiguities sufficiently small.
We show numerical tests in \cref{sec:data}. Furthermore, the extracted values for $\bg$  approach the one-loop perturbative values smoothly. All-in-all it is
sufficient to determine $\bg$ for the parameters listed in
\cref{tab:LCP}.

\begin{table}[htbp]
        \centering
        \begin{tabular}{cccclcccc}
                \toprule  
                $L/a$ & $\beta$ & $\kappa=\kappa_c$ &  $c_1$& $\quad c_2$ & $\hat c_2$\\
                \midrule
                12        & 4.3020      & 0.135998 &  0.02391(26) &  0.078(18) &  0.000191(45) \\    
                16        & 4.4662      & 0.135607   & 0.02049(22) &    0.108(15) &  0.000141(20)   \\
                20        & 4.5997      & 0.135289  &  0.01899(79)  &  0.043(61) &  0.000036(50)    \\
                24        & 4.7141      & 0.135024  &   0.01796(38) &   0.261(33) &  0.000144(18) \\
                \midrule
                24        & 4.9000      & 0.134601      & 0.01331(56) &  0.163(38) & $-$\\
                24        & 5.0671      & 0.134241  &    0.01166(48)  &   0.083(33) & $-$\\
                24    & 5.1719  & 0.134025  & 0.00932(43)  &   0.123(29) & $-$    \\
                24    & 6.0000  & 0.132575  &  0.00428(30)  &  0.053(21) & $-$    \\
                24    & 8.0000  & 0.130391  & 0.00134(12)  &0.0214(85)& $-$    \\
                24    & 16.000  & 0.127496  &  0.00009(\phantom{0}3) &   0.0016(21)& $-$    \\
                \bottomrule
        \end{tabular}
        \caption{Bare parameters for which $\bg$ has been computed as well as
        the fit coefficients $c_1,c_2$ for $\sigma(0.18)$ (cf.~eq.~(\ref{eq:fits}))
        and $\hat c_2=(a/L)^2 Z_{\rm m}^{-2}\,c_2$, where $Z_{\rm m}$ is the quark mass renormalization in the SF-scheme at the renormalization scale $\mu_{\rm dec}$
        (see \cref{subsec:lcp} for the details).}
        \label{tab:LCP}
\end{table}

\subsection{Simulations}
\label{sec:simulations}

We simulate QCD with a L\"uscher-Weisz gauge action~\cite{Luscher:1984xn} and three flavours of $\rmO(a)$ improved Wilson
quarks~\cite{Sheikholeslami:1985ij,Bulava:2013cta}.
To implement the strategy described above, simulations of QCD at zero quark mass or even at slightly negative masses are 
necessary. A finite volume with suitably chosen boundary conditions is indispensable to prevent negative or very small
eigenvalues of the Wilson Dirac operator. With Schr\"odinger Functional boundary conditions, massless 
simulations of our action  on our LCP were unproblematic~\cite{DallaBrida:2016kgh}. Here, in order to avoid chiral symmetry breaking by the boundaries,
we work on a hyper-torus. Choosing anti-periodic boundary conditions for the fermionic fields in all four directions induces a sufficiently large spectral gap.

The simulation setup is similar to previous simulations with $\Nf=3$ degenerate quarks. We use an even-odd preconditioned 
variant of the HMC algorithm~\cite{Duane:1987de} in which the three quarks are separated into a doublet and a 
third quark. The quark determinant for the doublet is factorized~\cite{Hasenbusch:2001ne} into three factors
$\det[\hat D^\dagger\hat D] = \det[\hat D^\dagger\hat D+\mu_2^2]\times \det[(\hat D^\dagger\hat D+\mu_1^2)(\hat D^\dagger\hat D+\mu_2^2)^{-1}] \times \det[(\hat D^\dagger\hat D)(\hat D^\dagger\hat D+\mu_1^2)^{-1}]$ and pseudo fermion fields
are introduced for each factor. Here $\hat D$ denotes the even-odd preconditioned clover Wilson operator. 
For the third quark it is required that all eigenvalues of $\hat D$ have real parts that are  
larger than some $r_a>0$. The operator in the determinant $\det[\hat D]=\det[\sqrt{\hat D^\dagger \hat D}]$ can then be approximated  
by a rational function of $\hat D^\dagger \hat D$~\cite{Kennedy:1998cu,Clark:2006fx}. The inexactness of this approximation is accounted for by including a stochastically 
determined reweighting factor in the observables. The molecular dynamics equations derived from the 
various contributions to the action are solved using multi level integration schemes with two levels~\cite{Sexton:1992nu}.  
The gauge action is integrated with the finest resolution while forces from different pseudo-fermion actions 
are evaluated less frequently. We use a modified version of \verb+openQCD-1.6+~\cite{Luscher:2012av} for our simulations
and refer to its documentation for further details. Simulation parameters 
as well as tests concerning the spectral gap and the ergodicity of the simulations are collected in
appendix~\ref{app:B}.

\section{Data analysis}
\label{sec:data}

\subsection{Estimates for the derivatives entering \cref{eq:bg}}

In principle the derivatives can be obtained from insertions 
of $\partial_p S$, where $p$ denotes the bare quark mass and the bare coupling. However, such measurements have large statistical errors. Instead, for a given $L/a$, we simulate at neighboring values in $g_0^2$ and $\mq$, fit the data as a function of $\mq$ at fixed $g^2_0$  and vice versa, and then obtain the derivative 
as the derivative of the fit functions.
In detail we fit
\begin{equation}
	\label{eq:fits}
	\langle{O_g}\rangle|_{L/a,g_{0}^2}=\sum_{k=0}^{n_m-1} c_{k} \,(am_{\rm q})^{k}\,,
	\qquad
	\langle{O_g}\rangle|_{L/a,m_{\rm q}=0}=\sum_{k=1}^{n_g} d_k\, g_0^{2k} \,,
\end{equation}
to the data in tables \ref{tab:g0der} and \ref{tab:m0der}
and mostly take the results with $n_m=n_g=3$ to three data points.  

\begin{figure}[htbp]
\centering
   \includegraphics[width=.49\linewidth]{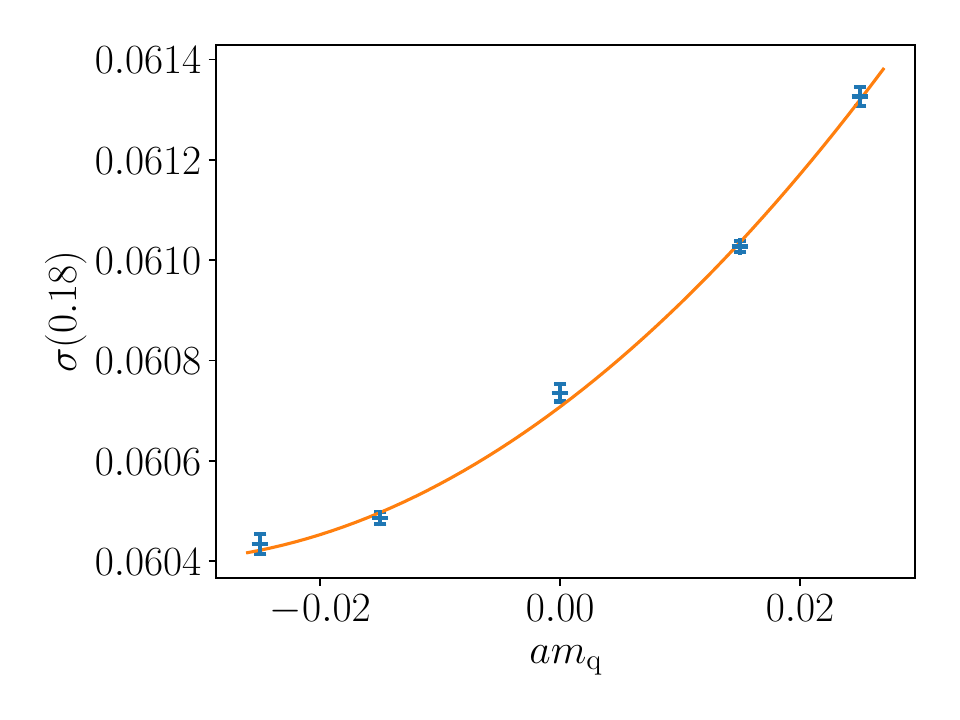}$\quad$\includegraphics[width=.49\linewidth]{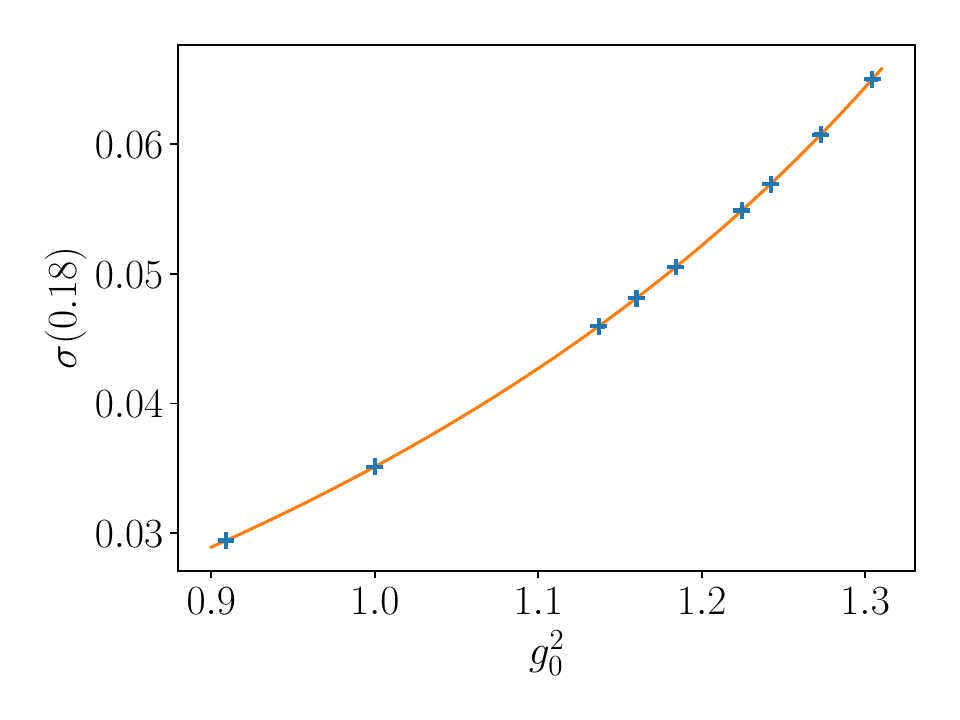}\\ 
   \caption{Quadratic fit of $\sigma(0.18)$ vs.~$a\mq$  for $L/a=24,\beta=4.7141$ (left), and nine-point fit with $n_g=5$ 
   	of $\sigma(0.18)$ vs. $g_0^2$ for $L/a=24$ and $4.6\leq \beta \leq 6.6$ (cf.~\cref{tab:g0der}) (right).}
   \label{fig:bgfit2}
\end{figure}	

Let us discuss this choice as well as the exceptions from it.
For the determination of the quark mass derivative the fitted data
are generically at $a\mq\approx 0,\pm0.025$. We need to check that this is small enough to obtain the first  derivative $\partial_{am_{\rm q}}\langle O_g \rangle\approx c_1 $, namely
that we are not affected by higher order terms in $a\mq$ beyond $n_m=3$. Indeed, note that 
the variable $a\mq$ in \cref{eq:fits} is natural for the linear term in $a\mq$ that we are seeking. However, already the 
quadratic term in the mass exists in the continuum limit. It is then naturally written in the form $\hat c_2 (L\mbar)^2  = c_2 (a\mq)^2$ with some renormalized quark mass $\mbar$ and a 
coefficient $\hat c_2(a/L)$ which has a finite limit $\hat c_2(0)$.
Therefore, as we increase $L/a$, the coefficient $c_2$ increases
in magnitude roughly at the rate $(L/a)^2$ (up to logarithmic renormalization effects). 
At $\beta=4.7141$ we simulated five quark mass points with $a\mq\approx0,\pm0.015,\pm0.025$. The corresponding data for $\sigma(0.18)$  are shown 
in \cref{fig:bgfit2}. One  sees  significant curvature, but also that a purely quadratic fit works  well. 
In \cref{tab:LCP} we list both $c_2$ and $\hat c_2 = (a/L)^2 Z_{\rm m}^{-2}\,c_2$ for the strict LCP.\footnote{We define the quark mass renormalization in the SF-scheme, $Z_\mathrm{m}=Z^{\rm SF}_\mathrm{m}= Z\,Z_\mathrm{A}/Z_\mathrm{P}$, at the renormalization scale
$\mu_\mathrm{dec}$, exactly as in \cite{DallaBrida:2022eua}, and also use the numerical values for $Z_\mathrm{A},Z_\mathrm{P}$ found in that reference.} Apart from the small statistics $L/a=20$ results, the expected behavior is clearly seen. It implies that the largest contamination of our data by
higher order effects in $a\mq$ is present for $L/a=24$.
Therefore we checked explicitly for $\beta=4.7141, L/a=24$, that we obtain the same derivative by fitting the available five points with $n_m=3$ and also by taking 
just the symmetric derivative, $\tilde\partial_{a\mq}f=[f(a\mq)-f(-a\mq)]/(2 a\mq)$ with either $a\mq =0.015$ or
$a\mq=0.025$. The resulting estimates for the derivative agree well
within errors:
\begin{gather}
	\tilde\partial_{am_{\rm q}} \sigma\big|_{am_{\rm q}= 0.025}=0.01786(54)\,,
	\qquad
	\tilde\partial_{am_{\rm q}} \sigma\big|_{am_{\rm q}= 0.015}=0.01806(55)\,,\\[1.5ex]
	\qquad
	\partial_{am_{\rm q}} \sigma\big|_{\rm 5pts}^\mathrm{fit}=0.01796(38) \,.
\end{gather}
Altogether we conclude that systematic errors due to this step are negligible
and use the five-point fit at $\beta=4.7141$ and the three-point fit otherwise.

The $g_0^2$ derivatives are obtained from the fits with $n_g=3$
to the data at the three closest values of $g_0^2$ of the data
compiled in \cref{tab:g0der}.
For the case of $L/a=24$, we compared these  with the estimates from $n_g=5$, fitting all nine available values of $g_0^2$
in the range specified by $4.6\leq \beta \leq 6.6$ (cf.~\cref{tab:g0der}).
The comparison, listed in \cref{tab:bgc018cmp} for $\sigma(0.18)$,
shows that the estimates for the $g_0^2$ derivative agree within uncertainties and the changes in the resulting $\bg$ are
negligible compared to the overall errors, which
are dominated by the ones of the quark mass derivative. The nine-point fit with $n_g=5$ is illustrated in \cref{fig:bgfit2}.

\begin{table}[htbp]
	\centering
	\begin{tabular}{cc  cc  cc }
		\toprule  
		$L/a$ & $\beta$ & \multicolumn{2}{c}{$g_0^2\partial \sigma/\partial g_0^2$ } & \multicolumn{2}{c}{$b_{\rm g}$} \\
		& & $n_g=5$ & $n_g=3$ & $n_g=5$ & $n_g=3$\\
		\midrule
		24 & 4.7141 &  0.16474(46) &  0.16531(58)  &  0.1090(23)  &  0.1086(24) \\ 
		24 & 4.9000 &  0.13907(20) &  0.1379(11)   &  0.0957(40)  &  0.0965(42) \\ 
		24 & 5.0671 &  0.12132(23) &  0.12195(63)  &  0.0961(40)  &  0.0957(40) \\ 
		24 & 5.1719 &  0.11211(20) &  0.11246(53)  &  0.0831(39)  &  0.0828(39) \\ 
		24 & 6.0000 &  0.06835(16) &  0.068680(86) &  0.0627(43)  &  0.0624(43) \\
		\bottomrule
	\end{tabular}
	\caption{Comparison of $b_{\rm g}$ and corresponding $g_0^2$ derivatives of $\sigma(0.18)$ from 3-point $n_g=3$ and 9-point $n_g=5$ fits, for $L/a=24$ and different values of $\beta$ (cf.~\cref{tab:g0der}).}
	\label{tab:bgc018cmp}
\end{table}

The data for $\hat\chi$ are analyzed in the same way. Results for $\bg$ from both observables are listed in \cref{tab:bg}.
As checks on the dependence on $c$, $R/L$, and $L/a$-values deviating from the lines of constant physics, there are additional data not reported in the tables. 

\begin{table}[htbp]
\centering
\begin{tabular}{ccccccc}
\toprule  
$L/a$ & $\beta$ & $\kappa$ & $g_0^2\partial\langle{O}_g\rangle/\partial g_0^2$ & $\partial\langle {O}_g\rangle/\partial am_{\rm q}$ & $b_{\rm g}$ \\
\toprule
12 & 4.3020 & 0.135998 & 0.17712(17) & 0.02392(26) & 0.1350(15) \\
& & & 1.1560(31) & 0.1661(47) & 0.1437(41) \\
16 & 4.4662 & 0.135607 & 0.16429(23) & 0.02049(22) & 0.1247(13) \\
& & & 1.1825(51) & 0.1518(48) & 0.1283(41) \\
20 & 4.5997 & 0.135289 & 0.1631(11) & 0.01898(79) & 0.1163(49) \\
& & & 1.269(24) & 0.167(19) & 0.132(15) \\
24 & 4.7141 & 0.135024 & 0.16531(58) & 0.01796(39) & 0.1086(24) \\
& & & 1.268(15) & 0.1398(95) & 0.1103(76) \\
24 & 4.9000 & 0.134601 & 0.1379(11) & 0.01331(56) & 0.0965(42) \\
& & & 1.085(31) & 0.099(14) & 0.091(13) \\
24 & 5.0671 & 0.134241 & 0.12195(63) & 0.01167(48) & 0.0957(40) \\
& & & 0.948(17) & 0.098(13) & 0.104(14) \\
24 & 5.1719 & 0.134025 & 0.11246(53) & 0.00931(43) & 0.0828(39) \\
& & & 0.871(15) & 0.078(11) & 0.090(13) \\
24 & 6.0000 & 0.132575 & 0.068680(86) & 0.00428(30) & 0.0624(43) \\
& & & 0.5497(25) & 0.0272(83) & 0.049(15) \\
24 & 8.0000 & 0.130391 & 0.033139(33) & 0.00134(12) & 0.0405(36) \\
& & & 0.2739(11) & 0.0126(37) & 0.046(14) \\
24 & 16.000 & 0.127496 & 0.0101876(67) & 0.000091(30) & 0.0089(30) \\
& & & 0.08898(21) & 0.00143(99) & 0.016(11) \\
\bottomrule
	\end{tabular}
	\caption{Results for $b_{\rm g}$, eq.~(\ref{eq:bg}), and corresponding derivatives w.r.t.~the bare parameters from $\langle O_g\rangle=\sigma(0.18)$  and $\langle O_g\rangle=\hat\chi$,
		in the first and second line for each value of $L/a$ and $\beta$, respectively.}
	\label{tab:bg}
\end{table}

\subsection{Relaxation of LCP condition for $\beta\geq 4.9$}

For our application to decoupling~\cite{DallaBrida:2022eua}, we 
also need bare couplings below $g_0^2=6/4.7$. Strictly following the
LCP  then requires prohibitively large $L/a$, in particular as we
seek the connection to perturbation theory below $g_0^2=1$.
We therefore keep $L/a=24$ fixed for all $\beta\geq 4.7$. 
In principle this means that there is an ambiguity of order $a/L=1/24$ 
which does not vanish as we increase $\beta$ and therefore decrease $a$. 
We follow  a skewed trajectory where $L$ decreases.

\begin{figure}[htbp]
	\centering
	\includegraphics[width=.49\linewidth]{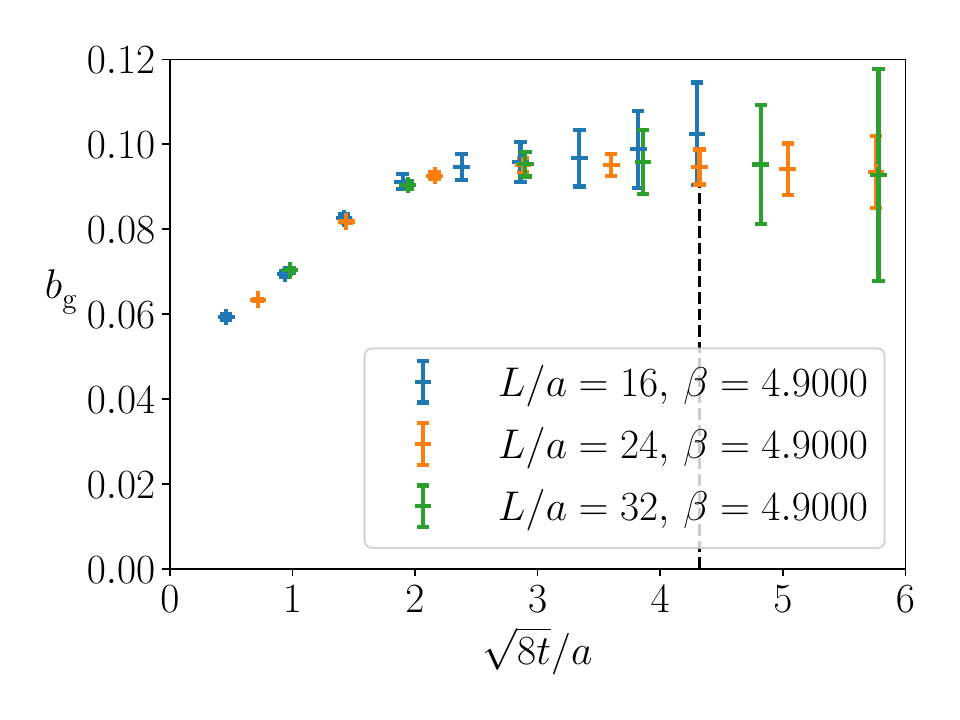} 
	\includegraphics[width=.49\linewidth]{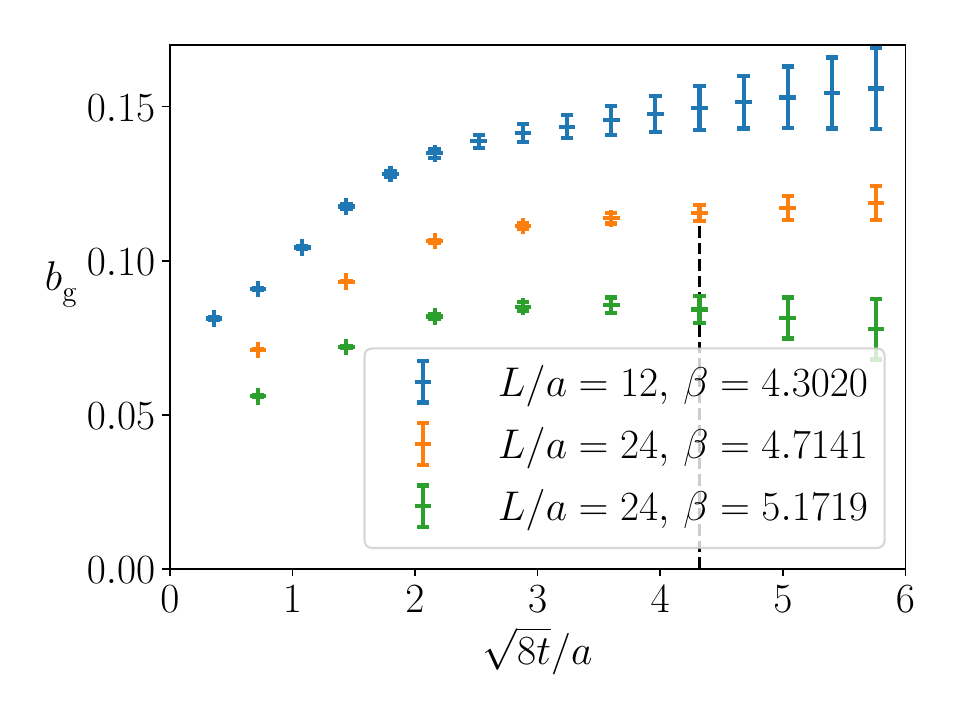}\\ 
	\caption{$\bg$, \cref{eq:bg}, from $\langle O_g\rangle=\sigma$ vs.~$\sqrt{8t}/a$ at $\beta=4.9$ for $L/a=16$, $24$, and $32$ (left) and at $\beta=4.3020,\,4.7141,\,5.1719$ for $L/a=12$, $24$, and $24$ (right). The  vertical dashed line indicates $\sqrt{8t}/a=0.18\times 24$, as used for all $\beta\geq4.7$ in the determination of $\bg$.}
	\label{fig:aLbg}
\end{figure}	

\begin{figure}[htbp]
	\centering
	\includegraphics[width=.49\linewidth]{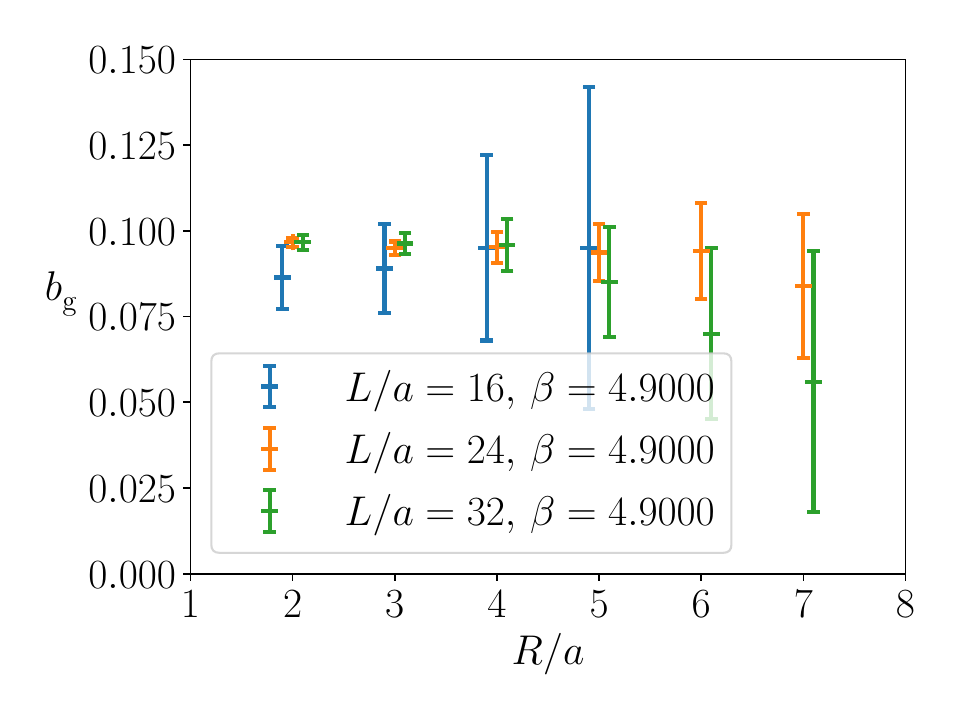} 
	\includegraphics[width=.49\linewidth]{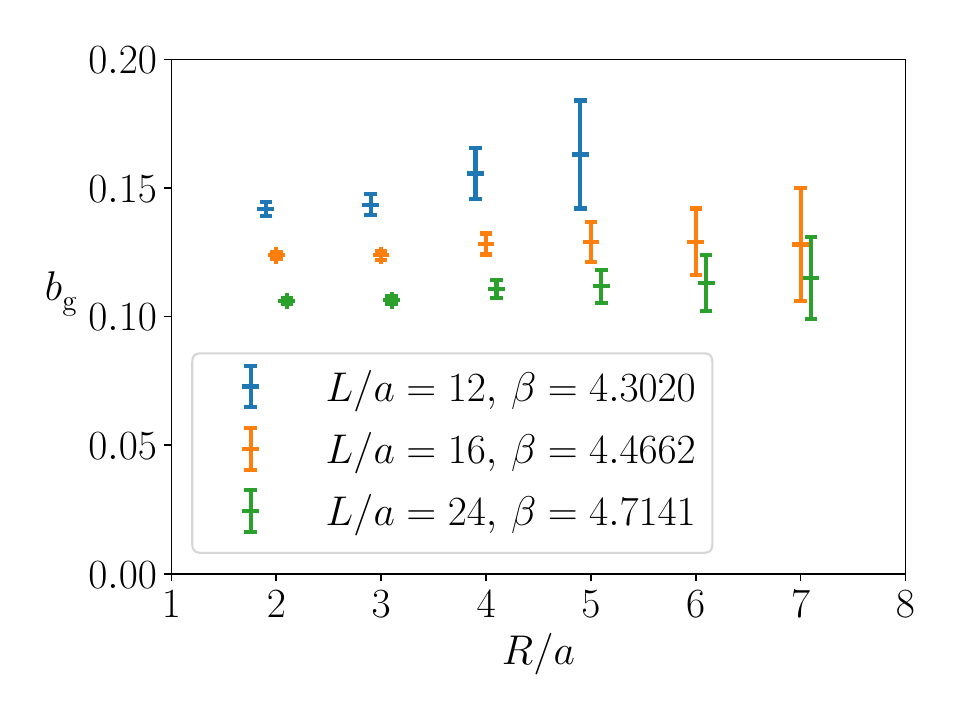}\\ 
	\caption{$\bg$, \cref{eq:bg}, from $\langle O_g\rangle=  R^2 \chi(R,R)$ vs.~$R/a$ at $\beta=4.9$ for $L/a=16$, $24$, and $32$ (left) and $\beta=4.3020$, $4.4662$, $4.7141$ for $L/a=12$, $16$, and $24$ (right).}
	\label{fig:aLbgchi}
\end{figure}

The magnitude of the ambiguity can be tested by 
computing how $\bg$ varies as one changes $L$. At fixed $a$, we therefore added computations with $L/a=16,32$. 
In \cref{fig:aLbg}, we show $\bg$ from $\sigma$, for $L/a=16,24,32$ and $\beta=4.9$. 
Within a small margin, a universal behavior, independent of $L/a$, is 
seen when $\bg$ is considered at fixed $\sqrt{8t}/a$. 
Therefore it is actually not very relevant that we deviate
from the fixed $L$ trajectory. Rather we have to be concerned 
about the  change in $t$ (which is linked to $L$ by $t=c^2L^2$ along the strict trajectory).
Fortunately, it is seen that 
$\bg$ becomes approximately independent of $t$ for $\sqrt{8t} \,\grtsim\, 3a$. Since $c=0.18$ puts us into the flat region for $L/a=24$, there are good indications that our deviation from the strict LCP is irrelevant at the level of our uncertainties. In \cref{fig:aLbgchi}, we show the same investigation 
for $\bg$ from $R^2 \chi(R,R)$, where the variable $\sqrt{8t}$ is 
replaced by $R$. The precision is not as good as in \cref{fig:aLbg}, but the statements made before hold,
exchanging $\sqrt{8t} \to R$.

In summary, we may relax the strict LCP as long as we keep 
$\sqrt{8t}$ and $R$ larger than about three in lattice units
and have $L/a\geq24$. 
The resulting ambiguity is small. We emphasize that 
above $g_0^2=6/4.9$, where the improvement by $\bg$ is most relevant,
we keep to the strict LCP.

\section{Results} 
\label{sec:results}

\begin{figure}[htbp]
\centering
\includegraphics[width=.49\linewidth]{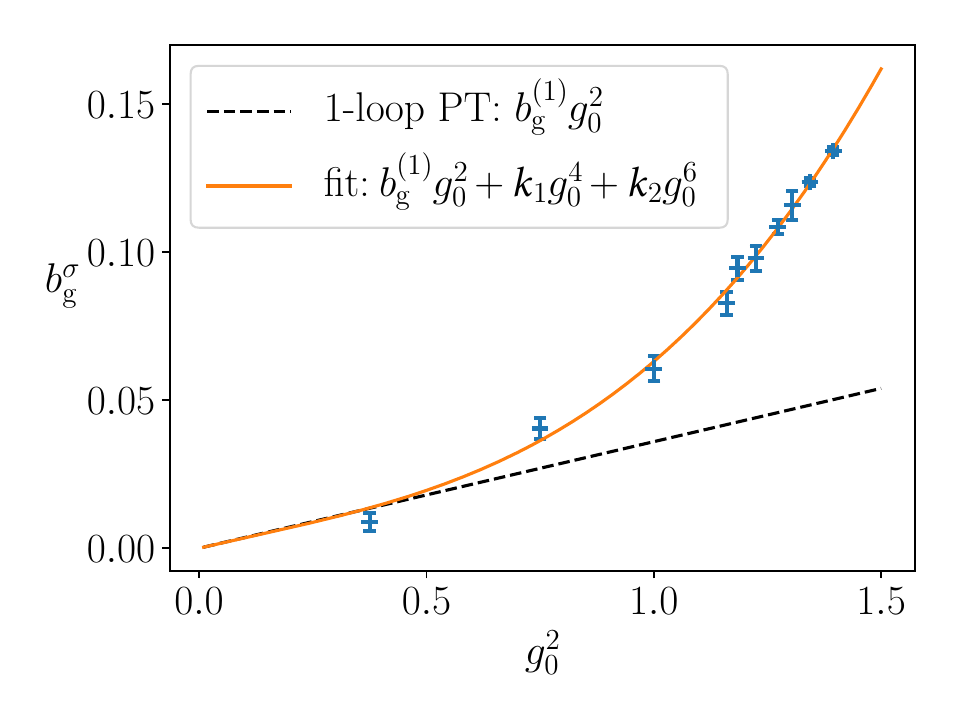}
\includegraphics[width=.49\linewidth]{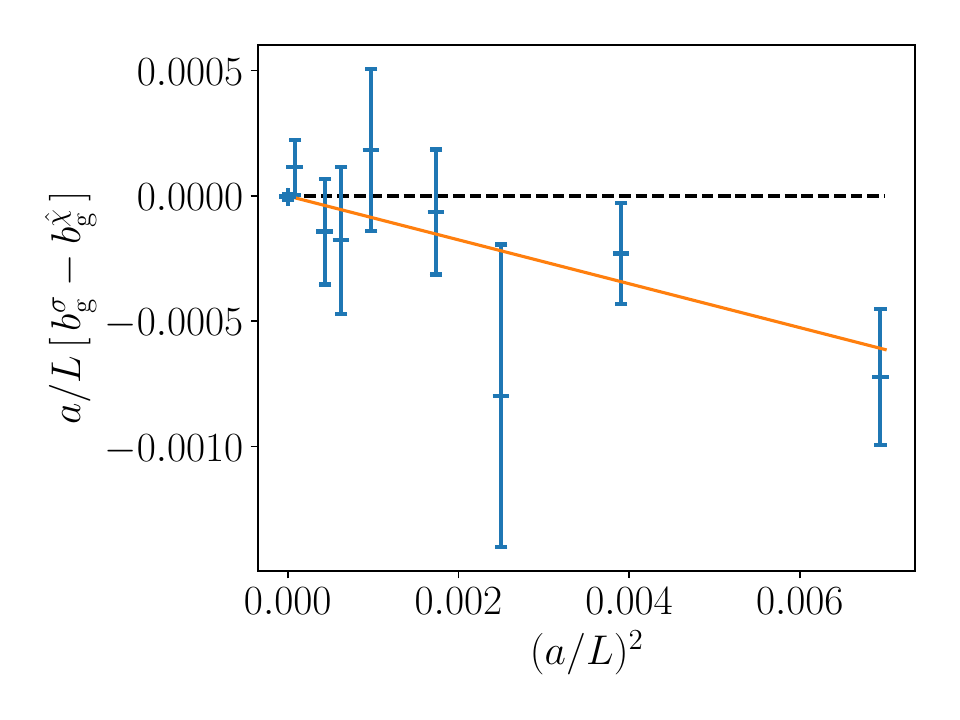}
\caption{Final result for $\bg$ from $\sigma(0.18)$ (left) and ambiguity $[\bg^\sigma-\bg^{\hat\chi}]\,\times \frac aL$ (right). The extra factor $a/L$ is added because $\bg$ always enters into observables with  an explicit factor of $a$; we also include a linear fit to all points constrained to go to zero. The values of $a/L$ considered in the plot have been obtained by fitting the values of $L/a$ as a function of $\beta_{\rm LCP}$ in Table 1 of ref.~\cite{DallaBrida:2022eua}, and enforcing the universal 2-loop running of the lattice spacing for $g_0^2\to0$.
}
	\label{fig:dbg}
\end{figure}

Our central results from the gradient flow energy density $\sigma(0.18)$ (cf.~\cref{tab:bg}) are displayed in \cref{fig:dbg}.
As an interpolation of the data from the perturbative regime up to 
our maximum bare coupling, 
we fit the results to
\begin{equation}
\bg^\sigma = g_0^2 \, \bigg[\bg^{(1)} +\sum_{i=1}^{n_k}k_i\,g_0^{2i}\bigg]\,, \quad \bg^{(1)}=0.036\,.
\label{eq:res1}
\end{equation}
A good fit is obtained already for $n_k=2$ with parameters
\begin{equation}
	k_1=-0.0151\,, 
	\quad
	k_2=0.0424\,. 
\label{eq:res2}
\end{equation}
This interpolation deviates little from others, obtained with e.g.~$n_k=3$ 
or from ones where we add a term proportional to $a/L$. The maximum (absolute) 
difference is $4\times10^{-4}$ in the important range of $4.30\leq \beta\leq 5.17$, and 
it is  $1.1\times10^{-3}$ overall.
Apart from this interpolation uncertainty, it is relevant to check for 
the always present ambiguity of $\rmO(a)$. We display the difference
to the determination from $\langle O_g\rangle=\hat\chi$ in \cref{fig:dbg}. 
The ambiguity is small and consistent with $\bg^\sigma-\bg^{\hat\chi}=\mathrm{const.}\times a/L$.
The uncertainty of the difference
is dominated by $\bg^{\hat\chi}$, but especially in the region of 
larger $g_0^2$, the test is rather significant and reassuring. 
The large difference of $\bg$ from the one-loop approximation is 
not due to our particular choice of improvement condition.

\section{Conclusions}
\label{sec:conclusions}

For the renormalization-by-decoupling strategy \cite{DallaBrida:2019mqg,DallaBrida:2022eua}, 
massive and massless QCD have to be connected with high precision. 
In such problems the improved bare coupling $\tilde{g}_0^2$ needs to be kept fixed as 
one approaches the continuum limit and the improvement coefficient $\bg$ is needed. 
The relevant combination $a\mq \bg$ reaches up to $0.06$ in the continuum extrapolation of the massive coupling 
in \cite{DallaBrida:2022eua}.
We have shown that $\bg$ can be determined requiring the absence of a term linear in
the quark mass around $\mq=0$ when boundary conditions do not
break chiral symmetry and one is not effectively in the large volume region where 
chiral symmetry is broken spontaneously.
We have further seen that with $L\approx 0.25\,\fm$, periodic boundary conditions for the gluon fields 
and anti-periodic ones for the quarks, the Dirac operator has a sufficiently large spectral gap such that the theory can be simulated
around the chiral limit. Furthermore, the quark mass derivative could be extracted  precisely and we obtained $\bg$ with good precision.
The result, \cref{eq:res1,eq:res2}, can be used without uncertainty, 
since subsequent assumptions made on the form of
the continuum limit will be more relevant than our errors. Still,
we have cited a systematic error which can be used if desired. 

We emphasize that \cref{eq:res1} has not been shown to be valid below $\beta=4.3$.
Firstly, we simulated only starting at that $\beta$-value, and secondly the LCP would quickly demand very 
small $\sqrt{8t}/a$ where ambiguities are potentially very large. Extending the present method to 
the region of CLS simulations~\cite{Bruno:2014jqa,cls:status} is therefore not entirely straightforward. 
One alternative are higher order perturbative estimates; for recent steps towards
a 2-loop result cf.~\cite{Costa:2023uba}, and one of us (MDB) has derived a preliminary result for the Wilson gauge action 
using the code developed for the 2-loop calculation of the SF coupling~\cite{Bode:1998hd,Bode:1999sm}. 
Another alternative is the chiral Ward identity technique, applied to the O($a$) improved 
flavour singlet scalar density, eq.~(\ref{eq:SRstructure}), cf.~\cite{Sint:1999ke}. 
The connection to $\bg$ relies on the identity, eq.~(\ref{eq:bg-ds}), which was 
first derived by Bhattacharya et al.~in \cite{Bhattacharya:2005rb}. In Section~\ref{sec:conditions}, we have pointed out
an oversight in this original derivation. The identity can however be rescued, provided one uses the particular 
discretization of the O($a$) counterterm to the scalar density, eq.~(\ref{eq:ds-counterterm}), 
which includes a contribution from the Sheikholeslami-Wohlert term.

\section*{Acknowledgements}

The work is supported by the German Research Foundation (DFG) research unit FOR5269 ``Future methods for studying confined gluons in QCD''.  
RH was supported by the programme ``Netzwerke 2021'', an initiative of the Ministry of Culture and Science of the State of Northrhine Westphalia, in the NRW-FAIR network, funding code NW21-024-A. SS and RS acknowledge funding by the H2020 program in the  {\em Europlex} training network, grant agreement No. 813942. 
Generous computing resources were supplied by the North-German Supercomputing Alliance (HLRN, project bep00072) and by the John von Neumann Institute for Computing (NIC) at DESY, Zeuthen.
The authors are grateful for the hospitality extended to them at CERN during the initial stage of this project. 
MDB would like to thank the members of the Lattice QCD group at Instituto de F\'isica Te\'orica of Madrid for their kind hospitality and support
during the final stages of this work.
We thank our colleagues in the ALPHA-collaboration, and especially Alberto Ramos, for valuable discussions. The sole responsibility for the content of this publication lies with the authors.

\begin{appendices}

\section{Derivation of Eqs.~(\ref{eq:bg-ds}) and (\ref{eq:bg})}
\label{app:A}
In this appendix we first derive the relation~Eq.~(\ref{eq:bg-ds}) using renormalized mass insertions
into the expectation value of a gradient flow observable. The gradient flow
makes this derivation transparent, as the on-shell condition for O($a$) improvement is never violated. 
Equation~(\ref{eq:bg}), which is key to our $\bg$-determination, is then easily established as
a corollary.

\subsection{Derivation of Eq.~(\ref{eq:bg-ds})}

At finite lattice spacing $a$ we consider some gluonic gradient flow observable, $\langle O_g(t)\rangle$, which is renormalized and
O($a$) improved once expressed in terms of the renormalized and O($a$) improved lattice QCD parameters~\cite{Luscher:2011bx}. 
Introducing the latter through
\begin{equation}
   \mr = Z_{\rm m}(\g02t,a\mu)\mqt,\qquad \gr^2= Z_{\rm g}(\g02t,a\mu)\g02t,
\end{equation}
we can {\em define} the insertion of the renormalized and O($a$) improved flavour singlet scalar density, $S_{\mathrm R}$, 
by taking the mass-derivative of the gradient flow observable,
\begin{equation}
 a^4 \sum_x \langle O_g(t) S^0_{\rm R}(x)\rangle_{\rm con} 
 \equiv a^4 \sum_x\left\{\langle O_g(t) S^0_{\rm R}(x)\rangle - \langle O_g(t)\rangle\langle S^0_{\rm R}(x)\rangle\right\}
 = -\left.\dfrac{\partial \langle O_g(t) \rangle}{\partial\mr}\right\vert_{\gr^2}\,.
 \label{eq:renmass-insertion}
\end{equation}
At fixed lattice spacing (i.e.~fixed $\g02t$) we have, 
\begin{equation}
  a^4 \sum_x \langle O_g(t) S^0_{\rm R}(x)\rangle_{\rm con}  =  -Z^{-1}_{\rm m}(\g02t,a\mu)\dfrac{\partial \langle O_g(t) \rangle}{\partial\mqt}\,,
 \label{eq:SRdef}
\end{equation}
with the improved bare mass insertion given by,
\begin{equation}
   -\dfrac{\partial \langle O_g(t) \rangle}{\partial\mqt} = \left\langle O_g(t) \frac{\partial S}{\partial\mqt}\right\rangle_{\rm con} 
   \equiv \left\langle O_g(t) \frac{\partial S}{\partial\mqt}\right\rangle - \biggl\langle O_g(t)\biggr\rangle \left\langle \frac{\partial S}{\partial\mqt}\right\rangle\,.
\label{eq:dOdmqt}
\end{equation}
Note that the insertion of the scalar density into the gradient flow observable does not give rise to contact terms, despite the 
Euclidean space-time summation in Eq.~(\ref{eq:renmass-insertion}). This means that 
no O($a$) counterterms are needed in Eq.~(\ref{eq:renmass-insertion}) beyond those of 
the scalar
density, Eq.~(\ref{eq:SRstructure}). 
Furthermore, the cubic divergence cancels in the connected correlation function, so that we can ignore the counterterm $\propto c_{\rm S}$. 

We now use the chain rule to relate the derivative of the action with respect to O($a$) improved bare parameters (depending on $\bg$ and $\bm$) 
to the derivatives with respect to the bare parameters. One may think of this as a change of variables,
\begin{equation}
   (g_0^2,a\mq) \rightarrow (\g02t,a\mqt)\,,
\end{equation}
where the latter are functions of the unimproved bare parameters. As is customary in the physics literature, we do not 
use different function names. Straightforward application of the chain rule gives,
\begin{eqnarray}
 \left.\dfrac{\partial S}{\partial \mq}\right\vert_{g_0^2}  &=&  \left.\dfrac{\partial S}{\partial \mqt}\right\vert_{\g02t}  
 \left.\dfrac{\partial \mqt}{\partial \mq}\right\vert_{g_0^2} + 
   \left.\dfrac{\partial S}{\partial \g02t}\right\vert_{\mqt}  
   \left.\dfrac{\partial \g02t}{\partial \mq}\right\vert_{g_0^2}\,, \\
  \left.\dfrac{\partial S}{\partial g_0^2}\right\vert_{\mq}  &=&  \left.\dfrac{\partial S}{\partial \mqt}\right\vert_{\g02t}  
  \left.\dfrac{\partial \mqt}{\partial g_0^2}\right\vert_{\mq} + 
   \left.\dfrac{\partial S}{\partial \g02t}\right\vert_{\mqt}  
   \left.\dfrac{\partial \g02t}{\partial g_0^2}\right\vert_{\mq}. 
\end{eqnarray}
We thus obtain a $2\times 2$ system of equations,
\begin{equation}
  \begin{pmatrix} M_{11} & M_{12} \\ M_{21} & M_{22} \end{pmatrix} 
  \begin{pmatrix} X_1 \\ X_2 \end{pmatrix} = \begin{pmatrix} A_1\\ A_2\end{pmatrix}\,,
\end{equation}
in the unknowns $X_1 = \partial S/\partial \mqt$ and $X_2= \partial S/\g02t$. Working out the 
derivatives with respect to the unimproved bare parameters we further have,
\begin{eqnarray}
  M_{11} = \left.\dfrac{\partial \mqt}{\partial \mq}\right\vert_{g_0^2}  &=&  1 + 2\bm a\mq\,, \\ 
  M_{21} = \left.\dfrac{\partial \mqt}{\partial g_0^2}\right\vert_{\mq}  &=&  a\mq^2 \bm'\,,  \\
  M_{12} = \left.\dfrac{\partial \g02t}{\partial \mq}\right\vert_{g_0^2} &=&  a g_0^2 \bg\,,  \\
  M_{22} = \left.\dfrac{\partial \g02t}{\partial g_0^2}\right\vert_{\mq} &=&  1+(\bg + g_0^2\bg')a\mq \,,
\end{eqnarray}
where we use the prime notation for derivatives w.r.t.~$g_0^2$, e.g.~$\bg' \equiv d\bg/dg_0^2$. 
The derivatives of the lattice action are given by,
\begin{eqnarray}
 A_1 = \left.\dfrac{\partial S}{\partial \mq}\right\vert_{g_0^2} &=& 
    a^4 \sum_x \psibar(x)\psi(x)\,, \\
 A_2 = \left.\dfrac{\partial S}{\partial g_0^2}\right\vert_{\mq} &=& 
    a^4 \sum_x \psibar(x) \frac{ia}4\csw' \sigma_{\mu\nu}F_{\mu\nu}\psi(x)  - \dfrac{S_g}{g_0^2}\,.
\end{eqnarray}
Inverting the matrix $M$,
\begin{equation}
  M^{-1} = \dfrac{1}{\det M} \begin{pmatrix} M_{22} & -M_{12} \\ -M_{21} & M_{11} \end{pmatrix}\,,
\end{equation}
solving for $X_1$, and inserting the explicit expressions for the matrix elements, we obtain,
\begin{equation}
  \left.\dfrac{\partial S}{\partial \mqt}\right\vert_{\g02t}  = 
      \dfrac{(1+ (\bg + g_0^2\bg')a\mq) \left.\dfrac{\partial S}{\partial \mq}\right\vert_{g_0^2} 
              -ag_0^2\bg \left.\dfrac{\partial S}{\partial g_0^2}\right\vert_{\mq}} 
            { \left(1+ (\bg + g_0^2\bg')a\mq\right) (1+ 2\bm a\mq)  -a^2\mq^2 g_0^2\bg \bm'}\,. 
\end{equation}
Expanding this expression to first order in $a\mq$ and neglecting O($a^2$) terms we then get,
\begin{equation}
  \left.\dfrac{\partial S}{\partial \mqt}\right\vert_{\g02t}  = 
    (1-2\bm a\mq)\left(\left.\dfrac{\partial S}{\partial \mq}\right\vert_{g_0^2} 
   -ag_0^2\bg \left.\dfrac{\partial S}{\partial g_0^2}\right\vert_{\mq} \right)\,.
   \label{eq:impmass-ins}
\end{equation}
Upon insertion into the connected correlation function, Eq.~(\ref{eq:dOdmqt}),
and comparing with Eqs.~(\ref{eq:SRstructure},\ref{eq:SRdef}) we thus see that we can identify,
\begin{equation}
   d_{\rm S} = -\dfrac{\bg}{2g_0^2}\,, \qquad b_{{\rm S}^0} = -2\bm\,,\qquad Z_{{\rm S}^0} = Z_{\rm m}^{-1}\,,
\end{equation}
provided that the $d_{\rm S}$-counterterm is realized on the lattice in the particular form of Eq.~(\ref{eq:ds-counterterm}).
We conclude that it is this lattice discretized form of the O($a$) counterterm that 
one must use in order to exploit the identity $\bg = -2g_0^2 d_{\rm S}$.

\subsection{Derivation of Eq.~(\ref{eq:bg})}

We now have a closer look at  Eq.~(\ref{eq:impmass-ins}) in the chiral limit, $\mq=0$. According to the discussion in Sect.~\ref{sec:conditions}, 
we impose the additional requirement that terms linear in the quark mass are absent. We thus obtain, 
in terms of derivatives of expectation values,
\begin{equation}
 \left.\dfrac{\partial\langle O_g(t)\rangle}{\partial \mq}\right\vert_{g_0^2,\mq=0} 
   -ag_0^2\bg \left.\dfrac{\partial\langle O_g(t) \rangle}{\partial g_0^2}\right\vert_{\mq=0} =0\,.
\end{equation}
Solving for $\bg$ yields Eq.~(\ref{eq:bg}) for gradient flow observables. In order to generalize this result to 
any O($a$) improved  gluonic observable, $O_g$, we recall that the gradient flow simply removed the complication from contact terms
in the insertion of the O($a$) improved scalar density, thereby allowing us to relate $\bg$ to $d_{\rm S}$. 
However, to derive Eq.~(\ref{eq:bg}), we only need to know that derivatives of O($a$) improved expectation values 
with respect to the improved bare parameters are again O($a$) improved. Hence the very same steps as carried out above 
establish Eq.~(\ref{eq:bg}) for arbitrary O($a$) improved gluonic observables.

\section{Simulations}
\label{app:B}
In the following we use the notation and conventions of~\cite{Luscher:2012av}
and the openQCD documentation. The simulation setup is very similar for all runs. 
The trajectory length is always 2 MDUs. Fermionic forces are integrated with a
4th order Omelyan-Mryglod-Folk (OMF4)~\cite{OMELYAN2003272} integrator with 8 integration steps per trajectory.
The gauge force is evaluated five times more frequently, as every interval between fermionic force 
evaluations is integrated with 1 OMF4 step.

The factorization of the determinant of the doublet into three determinants, 
as described in section~\ref{sec:simulations}, has in all cases the masses $a \mu_1=0.1$ and $a \mu_2=1.5$.
The degrees and ranges of the rational approximations for the third quark's determinant vary and can be found in 
\tab{tab:simpar}. This table also shows the acceptance rates. 
For solving linear systems, either a
standard conjugate gradient solver, or a variant that solves systems with different mass shifts simultaneously, is used and
stopped when a relative residuum norm below $10^{-12}$ is reached.

\begin{figure}[h]
\includegraphics[width=\linewidth]{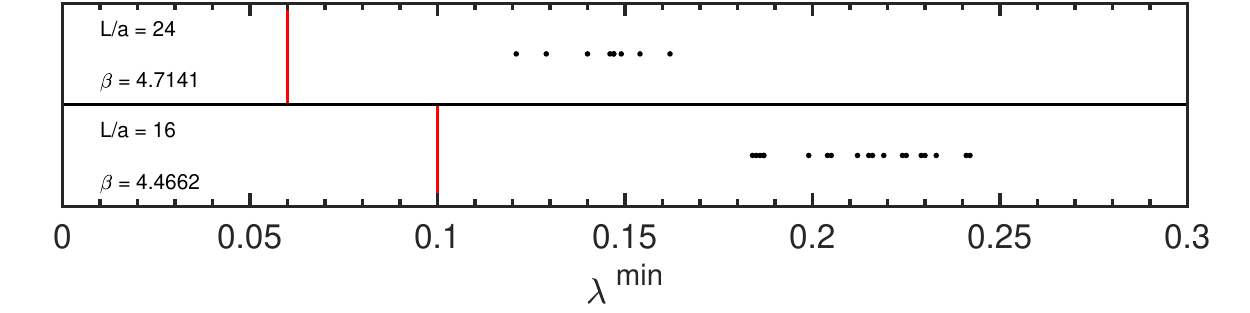}
\caption{The lowest eigenvalue of $|\gamma_5 \hat D|$ on several thermalized configurations at two different lattice spacings.
The upper part corresponds to the finer lattice spacing ($L/a=24, \beta=4.7141$), while the lower one to the coarser one ($L/a=16, \beta= 4.4662$). 
Both are at approximately zero quark mass. The vertical red lines correspond to the lower interval boundaries 
of the rational approximations that were used in the simulations.}\label{fig:ra}
\end{figure}

More elaborate solvers are not necessary, because the small volume together with our choice of boundary 
conditions results in quite well conditioned Dirac operators. We monitor the largest and the smallest 
eigenvalues of $|\gamma_5 \hat D|$ in order to decide on the ranges for the rational approximation, and in 
no case these were dangerously close to zero. Two examples are shown in~\Fig{fig:ra}.

Data analysis is performed using ADerrors.jl~\cite{Ramos:2020scv,Ramos:2018vgu} for the gradient flow energy density and pyerrors~\cite{Joswig:2022qfe} for the Wilson loops, keeping track of the correct errors and auto-correlations via the $\Gamma$-method~\cite{Wolff:2003sm,Schaefer:2010hu}. 

At last a remark on the ergodicity of the HMC algorithm. There are two potential difficulties with our setup.
First, it is known~\cite{Schaefer:2010hu} that with decreasing lattice spacing the correct sampling of topological sectors becomes
increasingly difficult. This does not affect our calculation because at the volume corresponding to our LCP
the sectors with non-zero charge are suppressed so strongly, that they play virtually no r\^ole. The second is a 
difficulty related to the center symmetry. In pure gauge theory, in the large volume limit, the center symmetry is spontaneously broken at high enough temperature. In our small volume
the symmetry is restored, but the trace of a Polyakov line exhibits a tri-modal distribution and 
many algorithms, including the HMC, tend to get stuck in one of the sectors. Although there is no 
center symmetry to break in the case of full QCD, similar algorithmic issues have been observed. We investigated 
the distribution of Polyakov lines and observed that at the volume corresponding to our LCP the distribution is 
unimodal, and in the perturbative region (e.g. $L/a=12, \beta=16$) there is enough movement between the ``sectors'' 
to ensure ergodicity.

\begin{table}[h!]
\centering
\begin{tabular}{l l l l l l l}
\toprule
$L/a$ & $\beta$ & $\kappa$ & degree & $r_a$  & $r_b$  & acceptance\\
\midrule
$ 12 $ & $ 4.1199 $ & $  0.136421625496234 $&$ 8 $&$ 0.1 $&$ 6.0 $&$ 98.8\%$\\
$ 12 $ & $ 4.302 $ & $  0.1350792062069652 $&$ 8 $&$ 0.1 $&$ 6.0 $&$ 98.9\%$\\
$ 12 $ & $ 4.302 $ & $  0.135997729474131 $&$ 8 $&$ 0.1 $&$ 6.0 $&$ 99.0\%$\\
$ 12 $ & $ 4.302 $ & $  0.1369288299729244 $&$ 8 $&$ 0.1 $&$ 6.0 $&$ 99.0\%$\\
$ 12 $ & $ 4.4662 $ & $  0.135602310355609 $&$ 8 $&$ 0.1 $&$ 6.0 $&$ 98.9\%$\\
$ 16 $ & $ 4.302 $ & $  0.136002904601631 $&$ 9 $&$ 0.1 $&$ 6.0 $&$ 98.2\%$\\
$ 16 $ & $ 4.4662 $ & $  0.1346938733548286 $&$ 9 $&$ 0.1 $&$ 6.0 $&$ 98.2\%$\\
$ 16 $ & $ 4.4662 $ & $  0.135607145941904 $&$ 9 $&$ 0.1 $&$ 6.0 $&$ 98.1\%$\\
$ 16 $ & $ 4.4662 $ & $  0.1365328877033368 $&$ 9 $&$ 0.1 $&$ 6.0 $&$ 98.0\%$\\
$ 16 $ & $ 4.6 $ & $  0.135286532423878 $&$ 9 $&$ 0.1 $&$ 6.0 $&$ 98.4\%$\\
$ 20 $ & $ 4.4662 $ & $  0.1356082 $&$ 10 $&$ 0.06 $&$ 6.0 $&$ 97.1\%$\\
$ 20 $ & $ 4.5997 $ & $  0.1343798945939137 $&$ 10 $&$ 0.06 $&$ 6.0 $&$ 96.7\%$\\
$ 20 $ & $ 4.5997 $ & $  0.1352889 $&$ 10 $&$ 0.06 $&$ 6.0 $&$ 98.0\%$\\
$ 20 $ & $ 4.5997 $ & $  0.1362102869948106 $&$ 10 $&$ 0.06 $&$ 6.0 $&$ 97.7\%$\\
$ 20 $ & $ 4.7141 $ & $  0.1350206 $&$ 10 $&$ 0.06 $&$ 6.0 $&$ 97.1\%$\\
$ 24 $ & $ 4.6 $ & $  0.135290 $&$ 10 $&$ 0.06 $&$ 6.0 $&$ 95.8\%$\\
$ 24 $ & $ 4.7141 $ & $  0.1341184114470437 $&$ 10 $&$ 0.06 $&$ 6.0 $&$ 95.9\%$\\
$ 24 $ & $ 4.7141 $ & $  0.1344791340035503 $&$ 10 $&$ 0.06 $&$ 6.0 $&$ 96.0\%$\\
$ 24 $ & $ 4.7141 $ & $  0.1350238708 $&$ 10 $&$ 0.06 $&$ 6.0 $&$ 95.7\%$\\
$ 24 $ & $ 4.7141 $ & $  0.1355730386938167 $&$ 10 $&$ 0.06 $&$ 6.0 $&$ 95.6\%$\\
$ 24 $ & $ 4.7141 $ & $  0.1359416391158157 $&$ 10 $&$ 0.06 $&$ 6.0 $&$ 95.5\%$\\
$ 24 $ & $ 4.83 $ & $  0.134758 $&$ 10 $&$ 0.06 $&$ 6.0 $&$ 96.5\%$\\
$ 24 $ & $ 4.9 $ & $  0.133700810592398 $&$ 10 $&$ 0.06 $&$ 6.0 $&$ 96.0\%$\\
$ 24 $ & $ 4.9 $ & $  0.134600621200435 $&$ 10 $&$ 0.06 $&$ 6.0 $&$ 95.9\%$\\
$ 24 $ & $ 4.9 $ & $  0.1355126253782558 $&$ 10 $&$ 0.06 $&$ 6.0 $&$ 95.9\%$\\
$ 24 $ & $ 5.0671 $ & $  0.1333461402387992 $&$ 10 $&$ 0.06 $&$ 6.0 $&$ 96.4\%$\\
$ 24 $ & $ 5.0671 $ & $  0.134241167314929 $&$ 10 $&$ 0.06 $&$ 6.0 $&$ 96.1\%$\\
$ 24 $ & $ 5.0671 $ & $  0.1351482905289904 $&$ 10 $&$ 0.06 $&$ 6.0 $&$ 96.1\%$\\
$ 24 $ & $ 5.1719 $ & $  0.1331332571760407 $&$ 10 $&$ 0.06 $&$ 6.0 $&$ 96.6\%$\\
$ 24 $ & $ 5.1719 $ & $  0.134025419206206 $&$ 10 $&$ 0.06 $&$ 6.0 $&$ 96.3\%$\\
$ 24 $ & $ 5.1719 $ & $  0.1349296191446659 $&$ 10 $&$ 0.06 $&$ 6.0 $&$ 95.9\%$\\
$ 24 $ & $ 5.2767 $ & $  0.133817232093422 $&$ 10 $&$ 0.06 $&$ 6.0 $&$ 96.1\%$\\
$ 24 $ & $ 6.0 $ & $  0.131701815057449 $&$ 10 $&$ 0.06 $&$ 6.0 $&$ 96.4\%$\\
$ 24 $ & $ 6.0 $ & $  0.132574832360087 $&$ 10 $&$ 0.06 $&$ 6.0 $&$ 95.8\%$\\
$ 24 $ & $ 6.0 $ & $  0.1334595009080744 $&$ 10 $&$ 0.06 $&$ 6.0 $&$ 95.7\%$\\
$ 24 $ & $ 6.6 $ & $  0.131762512060963 $&$ 10 $&$ 0.06 $&$ 6.0 $&$ 95.8\%$\\
$ 24 $ & $ 7.3 $ & $  0.131001797051939 $&$ 10 $&$ 0.06 $&$ 6.0 $&$ 94.6\%$\\
$ 24 $ & $ 8.0 $ & $  0.1295464517539626 $&$ 10 $&$ 0.06 $&$ 6.0 $&$ 94.7\%$\\
$ 24 $ & $ 8.0 $ & $  0.130391036560308 $&$ 10 $&$ 0.06 $&$ 6.0 $&$ 94.9\%$\\
$ 24 $ & $ 8.0 $ & $  0.1312467062640528 $&$ 10 $&$ 0.06 $&$ 6.0 $&$ 94.6\%$\\
$ 24 $ & $ 8.8 $ & $  0.129827754035436 $&$ 10 $&$ 0.06 $&$ 6.0 $&$ 94.0\%$\\
$ 24 $ & $ 13.5849 $ & $  0.127974784827748 $&$ 10 $&$ 0.06 $&$ 6.0 $&$ 89.0\%$\\
$ 24 $ & $ 16.0 $ & $  0.1266880690255924 $&$ 10 $&$ 0.06 $&$ 6.0 $&$ 83.6\%$\\
$ 24 $ & $ 16.0 $ & $  0.127495678088902 $&$ 10 $&$ 0.06 $&$ 6.0 $&$ 83.1\%$\\
$ 24 $ & $ 16.0 $ & $  0.1283136498788704 $&$ 10 $&$ 0.06 $&$ 6.0 $&$ 83.0\%$\\
$ 24 $ & $ 19.4595 $ & $  0.12702996843205 $&$ 10 $&$ 0.06 $&$ 6.0 $&$ 75.3\%$\\
\bottomrule
\end{tabular}
\caption{The degrees and ranges of the rational approximations used in the simulations. The last column is 
the rounded average acceptance rate.}\label{tab:simpar}
\end{table}

\section{Tables of simulation results}
\label{app:C}

In tables \ref{tab:g0der} and \ref{tab:m0der} we list the results of our simulations, 
which are used to obtain the derivatives of $\sigma(c=0.18)$ and $\hat{\chi}$ with respect to the quark mass and bare coupling.

\begin{table}[htb]
\centering
\begin{tabular}{cccccccc}
\toprule  
$L/a$ & $\beta$ & $\kappa$ & $N_{\rm rep}$ & $N_{\rm ms}$ & $\sigma(0.18)$ & $\hat\chi$ \\
\toprule
12 & 4.1199 & 0.136422 & 6 & 23880 & 0.076801(11) & 0.53993(20) \\
12 & 4.3020 & 0.135998 & 6 & 23760 & 0.0683942(93) & 0.48549(17) \\
12 & 4.4662 & 0.135602 & 6 & 23880 & 0.0622623(79) & 0.44520(14) \\
16 & 4.3020 & 0.136003 & 6 & 24120 & 0.069545(12) & 0.53586(25) \\
16 & 4.4662 & 0.135607 & 6 & 42840 & 0.0628847(77) & 0.48830(17) \\
16 & 4.6000 & 0.135287 & 6 & 24120 & 0.0583230(85) & 0.45526(20) \\
20 & 4.4662 & 0.135608 & 6 & 2292 & 0.066489(43) & 0.53472(95) \\
20 & 4.5997 & 0.135289 & 6 & 3360 & 0.061430(32) & 0.49651(79) \\
20 & 4.7141 & 0.135021 & 6 & 2244 & 0.057588(39) & 0.46590(80) \\
24 & 4.6000 & 0.135290 & 10 & 9600 & 0.065001(22) & 0.53076(54) \\
24 & 4.7141 & 0.135024 & 10 & 13800 & 0.060736(18) & 0.49761(42) \\
24 & 4.8300 & 0.134758 & 10 & 9600 & 0.056919(18) & 0.46874(49) \\
24 & 4.9000 & 0.134601 & 10 & 8200 & 0.054882(20) & 0.45275(52) \\
24 & 5.0671 & 0.134241 & 10 & 8200 & 0.050530(17) & 0.41831(46) \\
24 & 5.1719 & 0.134025 & 10 & 8440 & 0.048125(15) & 0.39982(39) \\
24 & 5.2767 & 0.133817 & 10 & 8800 & 0.045962(14) & 0.38294(39) \\
24 & 6.0000 & 0.132575 & 10 & 6600 & 0.035124(11) & 0.29651(29) \\
24 & 6.6000 & 0.131763 & 10 & 6600 & 0.0294367(84) & 0.25085(25) \\
24 & 7.3000 & 0.131002 & 10 & 9600 & 0.0247770(52) & 0.21282(17) \\
24 & 8.0000 & 0.130391 & 10 & 9600 & 0.0214180(44) & 0.18482(13) \\
24 & 8.8000 & 0.129828 & 10 & 9600 & 0.0185570(35) & 0.16133(11) \\
24 & 13.5849 & 0.127975 & 10 & 7800 & 0.0103696(21) & 0.091990(61) \\
24 & 16.000  & 0.127496 & 10 & 16800 & 0.0084930(11) & 0.075731(35) \\
24 & 19.4595 & 0.127030 & 10 & 7140 & 0.0067437(13) & 0.060316(43) \\
\bottomrule
	\end{tabular}
\caption{Results for $\sigma(0.18)$ and $\hat\chi$ used to obtain their 
 $g_0^2$-derivative. All measurements are carried out at vanishing quark mass and originate from  $N_{\rm rep}$ independent simulations with a total of $N_{\rm ms}$ measurements.}
\label{tab:g0der}
\end{table}

\begin{table}[htb]
\centering
\begin{tabular}{cccccccc}
\toprule  
$L/a$ & $\beta$ & $\kappa$ & $N_{\rm rep}$ & $N_{\rm ms}$ & $\sigma(0.18)$ & $\hat\chi$ \\
\toprule
12 & 4.3020 & 0.135079 & 6 & 23880 & 0.0690411(94) & 0.49040(17) \\
12 & 4.3020 & 0.136929 & 6 & 23880 & 0.0678453(92) & 0.48210(17) \\
16 & 4.4662 & 0.134694 & 6 & 42840 & 0.0634640(78) & 0.49311(17) \\
16 & 4.4662 & 0.136533 & 6 & 42840 & 0.0624397(76) & 0.48552(17) \\
20 & 4.5997 & 0.134380 & 6 & 4092 & 0.061932(29) & 0.50028(69) \\
20 & 4.5997 & 0.136210 & 6 & 4008 & 0.060983(27) & 0.49194(66) \\
24 & 4.7141 & 0.134118 & 10 & 13200 & 0.061326(19) & 0.50417(45) \\
24 & 4.7141 & 0.134479 & 10 & 33000 & 0.061027(12) & 0.50067(29) \\
24 & 4.7141 & 0.135573 & 10 & 33000 & 0.060485(12) & 0.49659(29) \\
24 & 4.7141 & 0.135942 & 10 & 13200 & 0.060433(20) & 0.49701(49) \\
24 & 4.9000 & 0.133701 & 10 & 8260 & 0.055317(21) & 0.45685(53) \\
24 & 4.9000 & 0.135513 & 10 & 7760 & 0.054652(19) & 0.45191(46) \\
24 & 5.0671 & 0.133346 & 10 & 8720 & 0.050873(17) & 0.42154(45) \\
24 & 5.0671 & 0.135148 & 10 & 8600 & 0.050290(17) & 0.41663(45) \\
24 & 5.1719 & 0.133133 & 10 & 8580 & 0.048435(16) & 0.40215(40) \\
24 & 5.1719 & 0.134930 & 10 & 8620 & 0.047969(15) & 0.39823(36) \\
24 & 6.0000 & 0.131702 & 10 & 6600 & 0.035265(11) & 0.29760(30) \\
24 & 6.0000 & 0.133460 & 10 & 6600 & 0.035050(10) & 0.29624(28) \\
24 & 8.0000 & 0.129546 & 10 & 9520 & 0.0214650(42) & 0.18502(13) \\
24 & 8.0000 & 0.131247 & 10 & 9600 & 0.0213978(43) & 0.18565(13) \\
24 & 16.000 & 0.126688 & 10 & 16800 & 0.0084963(11) & 0.075731(35) \\
24 & 16.000 & 0.128314 & 10 & 16800 & 0.0084918(11) & 0.075659(35) \\
\bottomrule
	\end{tabular}
\caption{Simulation parameters for the $\mq$-fits.
Results for $\sigma(0.18)$ and $\hat\chi$ used to obtain their 
 $\mq$-derivative. All measurements are carried out at  $a\mq\approx\pm 0.025$; for $\beta=4.7141$ also at $a\mq\approx\pm 0.015$. We carried out $N_{\rm rep}$ independent simulations for a total of $N_{\rm ms}$ measurements. 
 The measurements with $a\mq\approx0$ are listed in \cref{tab:g0der}.}
\label{tab:m0der}
\end{table}

\end{appendices}

\clearpage

\addcontentsline{toc}{section}{References}
\bibliographystyle{utphys}
\bibliography{paper}
\end{document}